\documentclass[10pt,twocolumn]{article}

\usepackage[T1]{fontenc}
\usepackage[utf8]{inputenc}
\usepackage[a4paper,
            top=25mm, bottom=25mm,
            left=18mm, right=18mm,
            columnsep=6mm]{geometry}

\usepackage{amsmath,amssymb}
\usepackage{graphicx}
\usepackage{xcolor}
\usepackage{booktabs}
\usepackage{tabularx}
\usepackage{multirow}
\usepackage{array}
\usepackage{longtable}
\usepackage{xurl}              % allows URL line-breaks in bibliography
\usepackage{hyperref}
\usepackage{cleveref}
\usepackage{enumitem}
\usepackage{titlesec}
\usepackage{fancyhdr}
\usepackage{abstract}
\usepackage{authblk}
\usepackage{caption}
\usepackage{subcaption}
\usepackage{float}
\usepackage{balance}       % balance last page columns
\usepackage{parskip}

\definecolor{medgray}{RGB}{200,200,200}

\hypersetup{
  colorlinks=true,
  linkcolor=black,
  citecolor=black,
  urlcolor=black,
  pdftitle={NPU Hardware Evaluation: A Comparative Study of Edge AI Inference Accelerators},
  pdfauthor={Davide Baltieri, Tobia Peruzzi},
}

\titleformat{\section}{\large\bfseries}{\thesection}{0.7em}{}[\vspace{-2pt}\color{medgray}\rule{\linewidth}{0.4pt}]
\titleformat{\subsection}{\normalsize\bfseries}{\thesubsection}{0.6em}{}
\titleformat{\subsubsection}{\normalsize\itshape}{\thesubsubsection}{0.5em}{}

\newcommand{\FAIL}{\textbf{FAIL}}
\newcommand{\tops}[1]{\mbox{#1\,TOPS}}
\newcolumntype{C}[1]{>{\centering\arraybackslash}p{#1}}
\newcolumntype{L}[1]{>{\raggedright\arraybackslash}p{#1}}
\newcolumntype{R}[1]{>{\raggedleft\arraybackslash}p{#1}}

\begin{document}

% ── Title block (full width) ──────────────────────────────
\twocolumn[{%
\begin{@twocolumnfalse}
  \begin{center}
    {\rule{\linewidth}{2pt}}\\[6pt]
    {\LARGE\bfseries NPU Hardware Evaluation v1.0}\\[4pt]
    {\large\itshape A Comparative Study of Edge AI Inference Accelerators}\\[8pt]
    {\rule{\linewidth}{0.8pt}}\\[8pt]
    \vspace{4mm}
    {\normalsize Davide Baltieri, Tobia Peruzzi}\\
    {\normalsize\itshape Covision Lab}\\
    {\normalsize Bressanone, BZ, IT}\\
    {\normalsize \{davide.baltieri,tobia.peruzzi\}@covisionlab.com}\\[2pt]
    \vspace{4mm}
  \end{center}

  \begin{abstract}
  \vspace{4mm}
AI inference in production settings is becoming the dominant cost line in enterprise AI.
The AI inference market is projected to grow from \$87B in 2024 to \$349B by 2032
(18.9\% CAGR)~\cite{ref1}. Neural Processing Units (NPUs), chips built specifically for AI inference, are emerging as a compelling alternative to GPU-only architectures,
with 35–70\% lower power consumption at comparable throughput~\cite{ref3}.

This white paper systematically evaluates ten edge AI inference accelerators across three
hardware categories: ASIC NPUs (Hailo-8, Hailo-10H, Axelera Metis, Axelera Europa, EdgeCortix Sakura~II),
SoC DSPs (SiMa MLSoC, Qualcomm QCS6490, QCS8550), and integrated NPUs (Intel Lunar Lake,
AMD XDNA2), benchmarked against an NVIDIA RTX A5000 with TensorRT as a production-grade
baseline. Twelve reference models spanning convolutional, mobile and transformer architectures
are used as a consistent benchmark suite. Results are analysed for throughput, latency,
model compatibility, power efficiency, SDK maturity and product lifecycle.
  \end{abstract}
  \vspace{4mm}

  % ── Table of Contents (top-level sections only) ──────────
  \vspace{2mm}
  {\rule{\linewidth}{0.4pt}}\\[4pt]
  {\small\bfseries Contents}\\[4pt]
  {\small
  \hyperref[sec:summary]{Summary: NPUs for Inference}
  \dotfill \pageref{sec:summary}\\[2pt]
  \hyperref[sec:overview]{1\quad Project Overview}
  \dotfill \pageref{sec:overview}\\[2pt]
  \hyperref[sec:specs]{2\quad Hardware Platform Specifications}
  \dotfill \pageref{sec:specs}\\[2pt]
  \hyperref[sec:optimisation]{3\quad Model Optimization Techniques}
  \dotfill \pageref{sec:optimisation}\\[2pt]
  \hyperref[sec:results]{4\quad Results and Analysis}
  \dotfill \pageref{sec:results}\\[2pt]
  }
  {\rule{\linewidth}{0.4pt}}
  \vspace{4mm}

\end{@twocolumnfalse}
}]

% ============================================================
\clearpage
\section*{Summary: NPUs for Inference}
\label{sec:summary}

\begin{table}[H]
\centering
\caption{Decision criteria for NPU adoption}
\small
\begin{tabularx}{\linewidth}{@{} L{2.2cm} X @{}}
\toprule
\textbf{Criterion} & \textbf{Key signal} \\
\midrule
TCO \& cost per inference & TCO priority jumped from 34\% to 41\% in a single quarter, overtaking raw performance as the dominant procurement lens.~\cite{ref2} \\
Power efficiency & NPU-based servers match or exceed GPU throughput while consuming 35--70\% less power; no liquid cooling required, cutting electricity costs by 60\%+.~\cite{ref3,ref5} \\
Data sovereignty \& compliance & On-device inference keeps data off cloud servers, critical under GDPR Art.~25, HIPAA, and DORA (EU banking, live Jan~2025).~\cite{ref6,ref7} \\
Latency for real-time workloads & Autonomous vehicles, industrial automation, and medical imaging require sub-10\,ms inference, which cloud round-trips cannot deliver.~\cite{ref8} \\
Reducing NVIDIA lock-in & With CUDA-based codebases as a strategic liability, firms qualify NPU alternatives to reduce single-vendor dependency.~\cite{ref4,ref9} \\
Software ecosystem maturity & Intel OpenVINO and ONNX Runtime are narrowing the CUDA gap; x86 compatibility remains the enterprise baseline (87\% market share, 2025).~\cite{ref10} \\
\bottomrule
\end{tabularx}
\end{table}

\subsection*{Key Findings}

\begin{table}[H]
\centering
\caption{Platform strengths and weaknesses}
\small
\begin{tabularx}{\linewidth}{@{} L{1.9cm} X @{}}
\toprule
\textbf{Platform} & \textbf{Strengths / Weaknesses} \\
\midrule
NVIDIA A5000 + TRT & Full model coverage, lowest single-stream latency, mature toolchain. High power (230\,W), high cost. \\
Qualcomm IQ-9075 & Best latency among DSPs, full coverage, good SDK. Sequential execution limits throughput. \\
Hailo-8 & Outstanding multi-stream throughput, excellent SDK usability. Limited operator support (4/12 models failed). \\
Axelera Metis & High peak TOPS, near-linear multi-chip scaling. Same operator coverage limits as Hailo-8. \\
EdgeCortix Sakura~II & Partially different operator support from Hailo/Axelera. BF16 support. Lower multi-stream throughput. \\
QCS6490 / QCS8550 & Full SoC integration, product longevity program. Sequential DSP limits throughput; fragmented SDK. \\
\bottomrule
\end{tabularx}
\end{table}

For full model flexibility with production-grade tooling, the \textbf{NVIDIA A5000} remains the reference. For embedded SoC deployments requiring long lifecycle support, \textbf{Qualcomm QCS8550 / IQ-9075} represent the strongest option. For workloads supported by the operator sets of ASIC NPUs, the \textbf{Hailo-8} and \textbf{Axelera Metis} offer the best performance per watt. 

\subsection*{Domain-Specific Applications}

\textbf{Healthcare.} Protected Health Information (PHI) must not leave the device. NPU-powered real-time diagnostic imaging is projected to grow to a \$2.2B segment by 2027.~\cite{ref11}

\textbf{Financial services.} DORA mandates sovereign audit rights; EU banks are actively repatriating inference workloads on-premise.~\cite{ref7}

\textbf{Automotive.} ADAS and sensor fusion require deterministic sub-10\,ms inference; 22M+ vehicles shipped with in-car NPUs in 2023 (22.8\% CAGR).~\cite{ref12}

\textbf{Manufacturing / IIoT.} Predictive maintenance and quality control AI; EU Data Act (Sept~2025) expands OT data sovereignty obligations.~\cite{ref13}

\textbf{Telecom.} 72\% of base stations integrate programmable NPUs for real-time 5G traffic management.~\cite{ref12}

\textbf{Enterprise IT.} Win~10 EOL refresh wave + Microsoft Copilot+ (40\,TOPS minimum) are forcing NPU decisions into standard PC procurement cycles.~\cite{ref14}

% ============================================================
\section{Project Overview}
\label{sec:overview}

\subsection{Objectives}

The primary objective is to systematically collect and consolidate technical know-how on a diverse set of hardware platforms for neural network inference at the edge. The evaluation addresses not only raw performance but also the practical aspects of working with each platform: toolchain quality, optimization workflow difficulty, model support breadth, and long-term hardware viability. The goal is to answer not only which device is fastest, but which device is most practical to adopt, maintain, and build upon.

A central finding that motivates the structure of this report is that peak hardware throughput is rarely the binding constraint in practice. Achieving production ready inference on dedicated accelerators (NPUs in particular) depends critically on how the model is prepared: reducing numerical precision through quantization, introducing weight sparsity, and removing redundant parameters through pruning. Each of these techniques interacts differently with each hardware platform; some accelerators expose dedicated INT4 or BF16 execution units, others require structured sparsity patterns to realize any speedup, and others impose operator constraints that make certain model families altogether incompatible. A fair hardware comparison therefore cannot treat model optimization as a post-hoc concern: it must be evaluated alongside the hardware itself. For this reason, Section~\ref{sec:optimisation} provides a grounded treatment of quantization, sparsification, and pruning before the benchmark results are presented.

Concretely, the same model can fail to compile on one NPU, run unoptimized on another, and approach peak hardware efficiency on a third, depending entirely on how it is quantized, sparsified, or pruned beforehand. Framing the optimization techniques explicitly in Section~\ref{sec:optimisation} allows the benchmark results that follow in Section~\ref{sec:results} to be read not merely as a leaderboard, but as a reflection of each platform's ability to exploit reduced-precision arithmetic and sparse or compact computation.

\subsection{Hardware Platforms}

Three categories of inference hardware were evaluated.

\textbf{ASIC NPUs} are dedicated silicon devices designed for neural network inference, implementing fixed, highly optimised dataflows and memory hierarchies. This specialisation enables exceptional performance-per-watt but at the cost of flexibility: supported operations, quantization formats, and network topologies are constrained by the fixed architecture. Evaluated devices: Hailo-8, Hailo-10H, Axelera Metis, Axelera Europa, and EdgeCortix Sakura~II.

\textbf{SoC DSPs} integrate a DSP or DSP array within a broader SoC alongside general-purpose CPU cores, memory controllers, and often additional accelerator blocks. DSPs offer greater programmability than pure ASIC solutions, and their integration within a full SoC makes them well-suited for complete embedded systems. Evaluated devices: SiMa MLSoC, Qualcomm QCS6490, and Qualcomm QCS8550.

\textbf{Integrated NPUs} are neural network acceleration blocks embedded directly within general-purpose consumer or workstation processors, sharing the same package and memory subsystem as the host CPU and GPU. Evaluated devices: Intel Lunar Lake (Core Ultra Series~2) and AMD XDNA2 (Ryzen AI 300 series).

As a performance reference baseline, an \textbf{NVIDIA RTX A5000} GPU was included, with all inference executed through TensorRT compiled in INT8 mode.

\subsection{Evaluation Methodology}

Each platform was assessed across the following dimensions:

\begin{enumerate}[noitemsep, topsep=2pt]
  \item \textbf{Installation and demo inference.} End-to-end hardware and software setup, producing a step-by-step installation guide and, where applicable, a reproducible container environment.
  \item \textbf{SDK-based network import, optimization, and quantization.} Full workflow from a floating-point model to a deployable, optimized artefact.
  \item \textbf{Model zoo analysis.} Breadth, documentation quality, and practical utility of vendor-provided model repositories.
  \item \textbf{Optimization assessment on reference models.} Exercise of the full optimization and quantization pipeline on twelve reference architectures.
  \item \textbf{Inference application development.} A minimal but complete inference application using the vendor's native SDK: model loading, raw tensor input (blob in), raw tensor output (blob out), and FPS measurement.
  \item \textbf{Performance analysis.} Throughput (FPS) relative to rated peak performance (TOPS) to derive an efficiency metric.
  \item \textbf{Product lifecycle.} Announced/expected product lifetime, SDK maturity and update frequency, supply chain continuity.
\end{enumerate}

\subsection{Reference Models}

A fixed set of twelve neural network models was selected to cover the major architectural families in modern computer vision (Table~\ref{tab:models}).

\begin{table}[H]
\centering
\caption{Reference benchmark models}
\label{tab:models}
\small
\begin{tabularx}{\linewidth}{@{} L{2.6cm} L{2.0cm} R{1.1cm} @{}}
\toprule
\textbf{Model} & \textbf{Family} & \textbf{Params} \\
\midrule
ResNet-18        & ResNet        & 11.69M  \\
ResNet-34        & ResNet        & 21.80M  \\
ResNet-50        & ResNet        & 25.56M  \\
ResNet-101       & ResNet        & 44.57M  \\
ResNeXt-101 32$\times$8d & ResNeXt   & 88.79M  \\
EfficientNet-B0  & EfficientNet  & 5.29M   \\
MobileNetV3-S-100 & MobileNet   & 2.54M   \\
MobileNetV4-Conv-S & MobileNet  & 3.77M   \\
MNASNet-0.75     & MNASNet       & 2.91M   \\
ConvNeXt-Tiny    & ConvNeXt      & 28.59M  \\
ViT-Tiny p16/224 & ViT           & 5.72M   \\
Tiny-ViT-5M/224  & Hybrid ViT    & 5.39M   \\
\bottomrule
\end{tabularx}
\end{table}

All models operate on $224\times224$ RGB inputs and are drawn from established, publicly available model repositories, primarily sourced from the PyTorch Image Models (\texttt{timm}) library.~\cite{ref15}

\newpage

% ============================================================
\section{Hardware Platform Specifications}
\label{sec:specs}

\textbf{Note on platform coverage.}
All platforms described in this section are included for completeness and situational awareness.
However, benchmarking results (Section~\ref{sec:results}) are available only for the platforms that
have been fully tested to date: Hailo-8, Axelera Metis, EdgeCortix Sakura~II, and the Qualcomm DSP
family (QCS6490, QCS8550, IQ-9075). The \textbf{Hailo-10H}, \textbf{Intel Lunar Lake}, and
\textbf{AMD XDNA2} platforms are currently under evaluation; benchmark data for these devices will
be reported in a future revision of this document. The \textbf{Axelera Europa} was not yet available
for hardware evaluation at the time of writing; benchmarking will be conducted once hardware access
is confirmed. The \textbf{SiMa MLSoC} was not benchmarked due to SDK integration constraints
described in Section~\ref{sec:specs}; evaluation remains planned pending improved software support.

\subsection{ASIC NPUs}

\subsubsection{Hailo-8}

The Hailo-8 delivers up to \tops{26} at a typical power consumption of 2.5\,W, making it one of the most power-efficient ASIC NPUs at the time of its release. A defining architectural feature is its fully integrated on-chip memory, which eliminates the need for external DRAM entirely. The chip is offered both as an M.2 module, exposing a PCIe Gen~3.0 interface (2 lanes in B+M key, 4 lanes in M-key), and as a HHHL PCIe card for server and industrial-PC deployments. Supported frameworks include TensorFlow, PyTorch, ONNX and Keras.~\cite{ref17}

\subsubsection{Hailo-10H}

The Hailo-10H delivers \tops{40} of INT4 performance (equivalent to \tops{20} INT8) with exceptional power efficiency. Unlike the Hailo-8, it includes a direct DDR interface for large models (LLMs, VLMs, Stable Diffusion). The module carries 8\,GB of LPDDR4 on-module memory and connects via a 4-lane PCIe Gen~3 interface, available in M.2 Key M (2242/2280) form factor as well as a HHHL PCIe card variant, with typical chip power below 3.5\,W.~\cite{ref17}

\textit{\textbf{Evaluation status:} The Hailo-10H has not yet been benchmarked. Hardware and SDK setup is in progress; results will be included in a future revision.}

\subsubsection{Axelera Metis}

The Axelera Metis is built around a proprietary RISC-V-based architecture employing Digital In-Memory Computing (D-IMC), delivering up to \tops{214} of INT8 performance at 15\,TOPS/W. It is offered both as a single-slot HHHL PCIe Gen~3 x4 card and as an M.2 2280 module, available in 4\,GB and 16\,GB DRAM configurations. The Voyager SDK supports TensorFlow, PyTorch, and ONNX.~\cite{ref18}

\subsubsection{EdgeCortix Sakura II}

The Sakura~II is built on EdgeCortix's second-generation Dynamic Neural Accelerator (DNA-II) architecture, delivering \tops{60} of INT8 and 30\,TFLOPS of BF16 within a typical power envelope of 8\,W. It is available both as a HHHL PCIe Gen~3 (x8/x16) card and as an M.2 2280 module. The memory subsystem uses a dual 64-bit channel LPDDR4x interface with up to 16\,GB on-board (peak bandwidth 68\,GB/s), complemented by 20\,MB of on-chip SRAM. The MERA software suite supports PyTorch, TensorFlow Lite, and ONNX.~\cite{ref19}

\subsubsection{Axelera Europa}

The Axelera Europa is the second-generation AIPU from Axelera AI, delivering \tops{629} of INT8 performance within a 45\,W TDP. The chip integrates eight second-generation AIPU cores alongside 16 RISC-V vector processing units for on-chip pre- and post-processing, 128\,MB of L2 SRAM, and a memory subsystem delivering 200\,GB/s external DRAM bandwidth. Supported precision formats include INT4, INT8, and INT16. The Europa is supported by the same Voyager SDK used for the Metis, enabling models trained in PyTorch, TensorFlow, and ONNX to be compiled and deployed across the full Axelera platform without SDK changes.~\cite{ref16}

\textit{\textbf{Evaluation status:} The Axelera Europa is currently in early access; hardware was not yet available during the evaluation period covered by this report. Benchmarking will be conducted once hardware access is confirmed; results will be included in a future revision.}

\begin{table*}[t]
\centering
\caption{ASIC NPU platform specifications}
\label{tab:asic_specs}
\small
\begin{tabularx}{\textwidth}{@{} L{2.0cm} X L{1.4cm} L{2.8cm} L{2.8cm} L{2.2cm} @{}}
\toprule
\textbf{Device} & \textbf{Peak TOPS} & \textbf{Power} & \textbf{Memory} & \textbf{Interface} & \textbf{Precision} \\
\midrule
Hailo-8        & 26 (INT8)                      & 2.5\,W     & None (on-chip)           & PCIe Gen~3 x2/x4   & INT8 \\
Hailo-10H      & 40 (INT4) / 20 (INT8)          & $<$3.5\,W  & 8\,GB LPDDR4             & PCIe Gen~3 x4      & INT4, INT8, FP16 \\
Axelera Metis  & 214 (INT8)                     & 8--15\,W   & 4--16\,GB DRAM           & PCIe Gen~3 x4      & INT8 \\
Axelera Europa & 629 (INT8)                     & 45\,W      & 128\,MB L2; 200\,GB/s BW & PCIe (TBD)         & INT4, INT8, INT16 \\
Sakura~II      & 60 (INT8) / 30\,TFLOPS (BF16) & 8\,W       & 16\,GB LPDDR4x           & PCIe Gen~3 x8/x16  & INT8, BF16 \\
\bottomrule
\end{tabularx}
\end{table*}

\subsection{SoC DSPs}

\subsubsection{SiMa MLSoC}

The SiMa MLSoC integrates a \tops{50} machine learning accelerator, a quad-core Arm Cortex-A65 application processor, a Synopsys EV74 computer vision DSP, H.264/H.265 hardware codec, 4\,MB on-chip memory, a 32-bit quad-channel LPDDR4 memory controller, and 8 lanes of PCIe Gen~4, all on a 16\,nm TSMC process node. The Palette software provides a unified compilation and deployment stack.~\cite{ref20}

\textit{\textbf{Evaluation status:} The SiMa MLSoC has not been benchmarked in this evaluation. At the time of testing, SiMa's inference SDK exclusively exposed a GStreamer-based pipeline interface, which is incompatible with our tensor-level benchmarking framework (raw blob in / blob out). Evaluation will be revisited once the SDK offers a direct tensor I/O path; results will be included in a future revision.}

\subsubsection{Qualcomm QCS6490}

The QCS6490 is a 6\,nm SoC integrating an octa-core Qualcomm Kryo 670 CPU, an Adreno 643 GPU, and a Qualcomm Hexagon AI Engine (DSP + HVX + Hexagon Tensor accelerator) delivering up to \tops{12}. It is part of Qualcomm's Product Longevity Program with supply commitment until July 2036. AI inference is accessed via the AI Engine Direct SDK and SNPE/QNN runtimes.~\cite{ref21}

\subsubsection{Qualcomm QCS8550}

The QCS8550 is Qualcomm's flagship IoT and embedded processor, built on 4\,nm, featuring an octa-core Kryo CPU (peak 3.36\,GHz), an Adreno 740 GPU, and an 8th-generation Hexagon AI Engine delivering \tops{48} of INT8 performance. Like the QCS6490, it is part of Qualcomm's Product Longevity Program.~\cite{ref21}

\subsubsection{Qualcomm Dragonwing IQ-9075}

The IQ-9075 is the entry point of Qualcomm's Dragonwing IQ9 industrial SoC series, pairing an octa-core Kryo Gen~6 (Cortex-A78C-based) CPU clocked at 2.1--2.36\,GHz with an Adreno~663 GPU and dual Hexagon Tensor Processors. AI throughput is offered in two SKU variants: 50 dense TOPS (QCS9075-AC) and 100 dense TOPS (QCS9075-AA), with sparse-computing configurations on some reference boards quoted up to 200\,TOPS@INT8. The platform supports up to 36\,GB of LPDDR5 (6$\times$16-bit channels @ 3200\,MHz) with inline ECC, and exposes two PCIe Gen~4 links (2-lane and 4-lane). Like the QCS6490 and QCS8550, the IQ-9075 is covered by Qualcomm's Product Longevity Program, with industrial-grade operation from $-40$ to $+115\,^\circ$C (junction) and a committed supply horizon exceeding 10 years. AI workloads are accessed via the AI Engine Direct SDK and QNN/SNPE runtimes, with framework support for TensorFlow, PyTorch, and ONNX.~\cite{ref25}

\begin{table*}[t]
\centering
\caption{SoC DSP platform specifications}
\label{tab:soc_specs}
\small
\begin{tabularx}{\textwidth}{@{} L{1.8cm} L{1.8cm} L{1.0cm} X X L{2.4cm} @{}}
\toprule
\textbf{Device} & \textbf{Peak AI TOPS} & \textbf{Process} & \textbf{CPU} & \textbf{AI Engine} & \textbf{Memory} \\
\midrule
SiMa MLSoC  & 50             & 16\,nm & 4$\times$ Cortex-A65 @ 1.15\,GHz      & ML Accel.\ + EV74 DSP                 & LPDDR4 (ext.) \\
QCS6490     & 12 (INT8)      & 6\,nm  & 8$\times$ Kryo 670                    & Hexagon DSP + HVX + Tensor           & LPDDR5 (ext.) \\
QCS8550     & 48 (INT8)      & 4\,nm  & 8$\times$ Kryo (peak 3.36\,GHz)       & 8th-gen Hexagon + Tensor             & LPDDR5 (ext.) \\
IQ-9075     & 50/100 (dense) & ---    & 8$\times$ Kryo Gen~6 @ 2.1--2.36\,GHz & Dual Hexagon HTP (HVX + HMX)         & Up to 36\,GB LPDDR5 \\
\bottomrule
\end{tabularx}
\end{table*}

\subsection{Integrated NPUs}

\subsubsection{Intel Lunar Lake (Core Ultra Series 2)}

The Intel Lunar Lake NPU4 architecture delivers up to \tops{48} NPU-alone, a threefold increase over the preceding Meteor Lake NPU3 generation. A key departure is the integration of on-package LPDDR5X memory (16\,GB or 32\,GB), reducing memory latency and cutting system power by $\approx$40\% versus off-package solutions. Total platform AI performance across CPU, GPU, and NPU reaches 120\,TOPS. Intel retained FP16 support, whereas competing solutions from AMD and Qualcomm top out at INT8. The NPU is accessible via the OpenVINO toolkit and the Windows AI platform (DirectML / ONNX Runtime).~\cite{ref22}

\textit{\textbf{Evaluation status:} The Intel Lunar Lake NPU has not yet been benchmarked. Hardware and SDK setup is in progress; results will be included in a future revision.}

\subsubsection{AMD XDNA2 (Ryzen AI 300 Series)}

The AMD XDNA2 NPU is a spatial dataflow architecture with 32 AI Engine tiles, delivering up to \tops{50} of INT8 performance, a claimed 5$\times$ compute and 2$\times$ power efficiency improvement over the first-generation XDNA. A notable addition is Block FP16 support, which performs computations at 8-bit integer throughput with 16-bit floating-point numerical accuracy. Off-chip memory bandwidth reaches up to 120\,GB/s (dual-channel LPDDR5); on-chip SRAM achieves 800\,GB/s for AI engine tiles. The NPU is accessible via AMD's Ryzen AI software stack and ONNX Runtime.~\cite{ref14}

\textit{\textbf{Evaluation status:} The AMD XDNA2 NPU has not yet been benchmarked. Hardware and SDK setup is in progress; results will be included in a future revision.}

\begin{table*}[t]
\centering
\caption{Integrated NPU platform specifications}
\label{tab:int_npu_specs}
\small
\begin{tabularx}{\textwidth}{@{} L{2.2cm} L{1.5cm} L{2.0cm} L{1.5cm} L{3.8cm} X @{}}
\toprule
\textbf{Device} & \textbf{NPU TOPS} & \textbf{Platform TOPS} & \textbf{Package TDP} & \textbf{Memory} & \textbf{Precision} \\
\midrule
Intel Lunar Lake & 48 & 120 & 17--30\,W & 16--32\,GB LPDDR5X (on-pkg) & INT8, FP16 \\
AMD XDNA2        & 50 & ---  & 28--54\,W & LPDDR5 (external)           & INT4, INT8, BF16, Block FP16 \\
\bottomrule
\end{tabularx}
\end{table*}

\subsection{Reference Baseline: NVIDIA RTX A5000}

The RTX A5000 (NVIDIA Ampere) features 8,192 CUDA cores, 256 third-generation Tensor Cores, and 24\,GB of GDDR6 ECC memory on a 384-bit bus (768\,GB/s bandwidth). Single-precision performance is 27.8\,TFLOPS; Tensor Core performance reaches 222.2\,TFLOPS (FP16 with sparsity). TDP is 230\,W via PCIe 4.0 x16. All inference in this evaluation is executed through TensorRT compiled in INT8 mode.~\cite{ref23}

\begin{table}[H]
\centering
\caption{NVIDIA RTX A5000 specifications}
\small
\begin{tabularx}{\linewidth}{@{} L{2.8cm} X @{}}
\toprule
\textbf{Specification} & \textbf{Value} \\
\midrule
CUDA Cores         & 8,192 \\
Tensor Cores       & 256 (3rd generation) \\
Tensor Performance & 222.2\,TFLOPS (FP16 w/ sparsity) \\
Memory             & 24\,GB GDDR6 ECC \\
Memory Bandwidth   & 768\,GB/s \\
TDP                & 230\,W \\
Host Interface     & PCIe 4.0 x16 \\
Inference Runtime  & TensorRT (INT8) \\
\bottomrule
\end{tabularx}
\end{table}

\subsection{Quick Reference Summary}

Table~\ref{tab:specs_summary} shows a summary of all the evaluated platforms in a side-by-side comparison. Qualcomm SoC power is system-dependent and not directly comparable to standalone accelerator TDP figures.

\begin{table*}[t]
\centering
\caption{Side-by-side specifications of all evaluated platforms.}
\label{tab:specs_summary}
\small
\begin{tabularx}{\textwidth}{@{} X L{2.2cm} R{3.2cm} L{2.4cm} L{3.6cm} L{2.6cm} @{}}
\toprule
\textbf{Device} & \textbf{Category} & \textbf{Peak TOPS} & \textbf{Typical Power} & \textbf{Memory} & \textbf{Interface} \\
\midrule
Hailo-8          & ASIC NPU      & 26 (INT8)             & 2.5\,W          & None (on-chip)             & PCIe Gen 3 x2/x4   \\
Hailo-10H        & ASIC NPU      & 40 (INT4) / 20 (INT8) & $<$3.5\,W       & 8\,GB LPDDR4              & PCIe Gen 3 x4      \\
Axelera Metis    & ASIC NPU      & 214 (INT8)            & 8--15\,W        & 4--16\,GB DRAM            & PCIe Gen 3 x4      \\
Axelera Europa   & ASIC NPU      & 629 (INT8)            & 45\,W           & 128\,MB L2; 200\,GB/s BW  & PCIe (TBD)         \\
Sakura II        & ASIC NPU      & 60 (INT8)             & 8\,W            & 16\,GB LPDDR4x            & PCIe Gen 3 x8/x16  \\
SiMa MLSoC       & SoC DSP       & 50                    & $\approx$5\,W & LPDDR4 (external)    & PCIe Gen 4 x8      \\
QCS6490          & SoC DSP       & 12 (INT8)             & SoC-level    & LPDDR5 (external)     & Integrated SoC     \\
QCS8550          & SoC DSP       & 48 (INT8)             & SoC-level    & LPDDR5 (external)     & Integrated SoC     \\
IQ-9075          & SoC DSP       & 50 / 100 (dense)      & SoC-level    & Up to 36\,GB LPDDR5    & PCIe Gen 4         \\
Intel Lunar Lake & Integrated NPU & 48 NPU / 120 platform & 17--30\,W (pkg) & 16--32\,GB LPDDR5X & Integrated SoC     \\
AMD XDNA2        & Integrated NPU & 50 (INT8)             & 28--54\,W (pkg) & LPDDR5 (external)  & Integrated SoC     \\
NVIDIA RTX A5000 & GPU (reference) & 222\,TFLOPS (tensor) & 230\,W       & 24\,GB GDDR6 ECC      & PCIe 4.0 x16       \\
\bottomrule
\end{tabularx}
\end{table*}

% ============================================================
\section{Model Optimization Techniques}
\label{sec:optimisation}

As deep learning models continue to scale in size and complexity, deploying them efficiently on dedicated hardware such as NPUs requires the application of model optimization techniques. Three complementary strategies bridge the gap between model accuracy and deployment efficiency: \emph{quantization}, \emph{sparsification}, and \emph{pruning}.

\subsection{Quantization}

Quantization reduces the numerical precision used to represent a model's weights and activations. Standard training operates in FP32; quantization maps these values to lower-precision formats (most commonly INT8, but increasingly INT4, INT3, or even INT2).

This is beneficial for two primary reasons: (1) lower-precision data types require less memory for storing model parameters and activations; (2) integer arithmetic is substantially cheaper than floating-point on most hardware. NPUs in particular are often optimized with dedicated integer execution units delivering dramatically higher throughput for quantized workloads.

\subsubsection{Quantization Hyperparameters}

\textbf{Static quantization} computes all scale factors and zero points once using a calibration dataset, achieving the best performance in throughput, latency, and memory. \textbf{Dynamic quantization} computes these on-the-fly during inference, yielding better accuracy but significantly reducing practical benefits.

Quantization can be \textbf{symmetric} (zero point fixed at zero) or \textbf{asymmetric} (unconstrained zero point). Asymmetric quantization better exploits the integer range when activation distributions are shifted, as after ReLU. It can also be applied \textbf{per-tensor} (single scale factor) or \textbf{per-channel} (separate scale per output channel).

\subsubsection{Calibration}

Calibration determines appropriate min-max ranges for each layer's outputs, defining scale factors and zero points. Methods include:
\begin{itemize}[noitemsep, topsep=2pt]
  \item \textit{MinMax}: uses observed minimum and maximum values.
  \item \textit{MovingAverageMinMax}: smooths min-max estimates via a running average.
  \item \textit{HistogramMSE}: minimises the MSE between FP32 and quantized outputs.
  \item \textit{HistogramEntropy}: minimises KL-divergence between distributions.
  \item \textit{HistogramPercentile}: clips outliers at a chosen percentile.
\end{itemize}
A poor calibration choice can cause accuracy drops as large as 10\%. There is no universally optimal method.

\subsubsection{Overflow Handling}

Mitigation strategies include saturation (compute at higher bitwidth then saturate-cast), overflow-aware calibration, overflow-aware QAT, and stochastic rounding (shown empirically to reduce quantization bias).

\subsubsection{Post-Training Quantization and QAT}

\textbf{Post-training quantization (PTQ)} converts an already-trained FP32 model to lower precision using a calibration dataset, without retraining. It can be enhanced with techniques such as AdaRound. \textbf{Quantization-aware training (QAT)} simulates quantization effects during training via straight-through estimators, allowing the model to learn representations robust to reduced precision. QAT is typically introduced gradually and followed by a final PTQ step.

\subsection{Sparsification}

Sparsification introduces zeros into weight matrices or activation tensors. By itself it offers limited benefits; its value lies primarily as a prerequisite for pruning or, when backed by hardware support, as a route to measurable inference speedups.

\subsubsection{Types of Sparsity}

Weights can be selected using \textbf{magnitude-based} (zeros out smallest-absolute-value weights; simple and efficient) or \textbf{gradient-based} criteria (selects by importance w.r.t.\ the loss; more informed but costlier). Sparsity structure varies from \textbf{unstructured} (individual weights, requires hardware support), \textbf{structured} (entire filters/channels, compatible with standard dense hardware), to \textbf{semi-structured} (regular blocks, a middle ground).

\subsubsection{Hardware Support for Sparsity}

\textbf{Block Floating Point (BFP)} represents weights in blocks sharing a common exponent, allowing the hardware to skip computation when a full block is zero. \textbf{N:M sparsity patterns} (e.g., 2:4 supported in NVIDIA Ampere) require exactly $N$ out of every $M$ consecutive weights to be non-zero, enabling compressed weight storage and doubled throughput through dedicated sparse execution units.~\cite{ref24}

\subsubsection{Enforcing Sparsity in Practice}

Sparsification is typically implemented by replacing standard layers with sparse-aware equivalents carrying a binary mask applied before each forward pass. The mask is updated after each training epoch, with sparsity introduced gradually until the target level is reached. Structured sparsity requires consistency across layer boundaries.

\subsection{Pruning}

Pruning permanently removes zeroed-out parameters or structural components. The result is a smaller, more compact architecture that requires fewer parameters and less computation, with benefits independent of hardware sparse execution support.

\subsubsection{Weight and Structured Pruning}

\textbf{Weight pruning} removes individual parameters based on a saliency criterion and fine-tunes the sparse model. \textbf{Structured pruning} eliminates entire neurons, convolutional filters, or attention heads driven to zero by structured sparsification. Because the pruned model retains a regular, dense layout, it is immediately usable with standard hardware without specialized sparse support.

Pruning entire layers is feasible only in residual networks, since a zero-weight layer can be removed as long as the residual connection preserves the signal path.

\subsubsection{Practical Pruning Process}

The most effective approach is an iterative cycle: train to convergence, prune a small percentage of the lowest-importance weights, retrain to recover accuracy, and repeat. There is no universal recipe; empirical experimentation remains necessary.

\subsection{Combined Application and Joint Optimization}

In practice the three techniques are rarely applied in isolation. Their interactions often offer synergies: pruning reduces model complexity, stabilizing subsequent quantization; quantization can push small residual weights to exactly zero, further increasing sparsity; a pruned, sparse model presents a simpler calibration landscape.

\textbf{Quantization-aware pruning} uses an importance score incorporating both weight magnitude and quantization error. \textbf{Iterative quantization and pruning} alternates between both steps during training. \textbf{Joint loss functions} incorporate both a quantization error term and a sparsity penalty directly into the training objective.

From a hardware evaluation perspective, NPUs differ significantly in how well they exploit quantized arithmetic, handle sparse computation, and benefit from structured versus semi-structured sparsity. The benchmarks presented in the following sections are designed to surface these differences.

% ============================================================
\section{Results and Analysis}
\label{sec:results}

This section presents benchmark results across all evaluated platforms, covering throughput and latency on the twelve reference models (Section~\ref{sec:overview}), multi-stream and multi-chip throughput scaling on the Hailo-8, Axelera Metis, and EdgeCortix Sakura~II, quantization strategy effects on the NVIDIA A5000, and a cross-platform discussion of key findings.

\subsection{Cross-Platform Throughput}
\label{sec:throughput}

Tables~\ref{tab:asic_results} and \ref{tab:soc_results} report latency (ms) and throughput (FPS) for all twelve reference models across ASIC NPUs, SoC DSPs, and the A5000 reference baseline. \FAIL{} indicates a model could not be compiled or executed due to unsupported operators or memory constraints.

For the Hailo-8, only single-stream throughput is shown in Table~\ref{tab:asic_results}; multi-stream pipelined throughput is reported separately in Table~\ref{tab:hailo_streaming}. No equivalent multi-stream scaling is available on the Qualcomm Hexagon DSP platforms, which execute one inference job at a time.

The most immediate finding is the significant variation in model compatibility across device categories. The NVIDIA A5000 with TensorRT successfully compiled and executed all twelve models. The Qualcomm DSP family achieved near-complete coverage (the sole exception being MobileNetV3 on QCS8550). The ASIC NPUs exhibited substantial compatibility gaps: Hailo-8 failed on four models (ConvNeXt-Tiny, EfficientNet-B0, MobileNetV3-Small, and Tiny-ViT-5M), Axelera Metis showed the same failure set plus ResNeXt-101 and ViT-Tiny, and EdgeCortix Sakura~II failed on ConvNeXt-Tiny, EfficientNet-B0, ViT-Tiny, and Tiny-ViT—though it succeeded on MobileNetV3 where both Hailo and Axelera failed. The consistent failure of all three ASIC NPUs on transformer-based models highlights a fundamental limitation of fixed-function silicon, with growing implications as model architectures diversify.

\begin{table*}[t]
\centering
\caption{ASIC NPU throughput vs.\ NVIDIA RTX A5000 (TensorRT INT8). \FAIL{} = model could not be compiled or run.}
\label{tab:asic_results}
\small
\begin{tabularx}{\textwidth}{@{} X r r r r r r r r @{}}
\toprule
& \multicolumn{2}{c}{\textbf{Hailo-8}} & \multicolumn{2}{c}{\textbf{Axelera Metis}} & \multicolumn{2}{c}{\textbf{EdgeCortix S2}} & \multicolumn{2}{c}{\textbf{A5000 TRT}} \\
\cmidrule(lr){2-3}\cmidrule(lr){4-5}\cmidrule(lr){6-7}\cmidrule(lr){8-9}
\textbf{Model} & \textbf{Lat.} & \textbf{FPS} & \textbf{Lat.} & \textbf{FPS} & \textbf{Lat.} & \textbf{FPS} & \textbf{Lat.} & \textbf{FPS} \\
\midrule
convnext\_tiny        & \FAIL & \FAIL & \FAIL  & \FAIL  & \FAIL  & \FAIL   & 0.87\,ms & 1218 \\
efficientnet\_b0      & \FAIL & \FAIL & \FAIL  & \FAIL  & \FAIL  & \FAIL   & 0.57\,ms & 1898 \\
mnasnet\_075          & 1.87\,ms &  3081 & 2.0\,ms &  1156 & 1.02\,ms &   956 & 0.27\,ms & 4481 \\
mobilenetv3\_small    & \FAIL & \FAIL & \FAIL  & \FAIL  & 0.83\,ms &  1162 & 0.34\,ms & 3393 \\
mobilenetv4\_conv\_s  & 1.28\,ms &  5948 & 1.5\,ms &  1581 & 0.76\,ms &  1262 & 0.24\,ms & 5299 \\
resnet18              & 1.20\,ms &  2509 & 1.8\,ms &  1131 & 1.09\,ms &   894 & 0.20\,ms & 6800 \\
resnet34              & 4.02\,ms &  1139 & 2.5\,ms &   580 & 1.56\,ms &   630 & 0.32\,ms & 3738 \\
resnet50              & 5.02\,ms &   771 & 2.9\,ms &   516 & 1.94\,ms &   507 & 0.40\,ms & 2844 \\
resnet101             & 7.17\,ms &   112 & 4.6\,ms &   330 & 2.79\,ms &   355 & 0.69\,ms & 1554 \\
resnext101\_32x8d     & 15.17\,ms &   61 & \FAIL  & \FAIL  & 14.32\,ms &    71 & 0.86\,ms & 1236 \\
tiny\_vit\_5m\_224    & \FAIL & \FAIL & \FAIL  & \FAIL  & \FAIL  & \FAIL   & 0.71\,ms & 1515 \\
vit\_tiny\_p16\_224   & 9.66\,ms &    94 & \FAIL  & \FAIL  & \FAIL  & \FAIL   & 0.48\,ms & 2317 \\
\bottomrule
\end{tabularx}
\end{table*}

\begin{table*}[t]
\centering
\caption{SoC DSP throughput vs.\ NVIDIA RTX A5000 (TensorRT INT8). \FAIL{} = model could not be compiled or run.}
\label{tab:soc_results}
\small
\begin{tabularx}{\textwidth}{@{} X r r r r r r r r @{}}
\toprule
& \multicolumn{2}{c}{\textbf{QCS6490}} & \multicolumn{2}{c}{\textbf{QCS8550}} & \multicolumn{2}{c}{\textbf{IQ-9075}} & \multicolumn{2}{c}{\textbf{A5000 TRT}} \\
\cmidrule(lr){2-3}\cmidrule(lr){4-5}\cmidrule(lr){6-7}\cmidrule(lr){8-9}
\textbf{Model} & \textbf{Lat.} & \textbf{FPS} & \textbf{Lat.} & \textbf{FPS} & \textbf{Lat.} & \textbf{FPS} & \textbf{Lat.} & \textbf{FPS} \\
\midrule
convnext\_tiny       & 6.40\,ms & 150 & 4.59\,ms & 450 & 1.57\,ms &  645 & 0.87\,ms & 1218 \\
efficientnet\_b0     & 2.30\,ms & 305 & 3.44\,ms & 720 & 0.69\,ms & 1503 & 0.57\,ms & 1898 \\
mnasnet\_075         & 1.30\,ms & 398 & 1.82\,ms & 928 & 0.33\,ms & 3377 & 0.27\,ms & 4481 \\
mobilenetv3\_small   & 1.20\,ms & 446 & \FAIL    & \FAIL & 0.31\,ms & 3523 & 0.34\,ms & 3393 \\
mobilenetv4\_conv\_s & 1.00\,ms & 444 & 1.49\,ms & 951 & 0.27\,ms & 4092 & 0.24\,ms & 5299 \\
resnet18             & 2.30\,ms & 344 & 1.48\,ms & 954 & 0.50\,ms & 2122 & 0.20\,ms & 6800 \\
resnet34             & 3.20\,ms & 261 & 1.84\,ms & 795 & 0.76\,ms & 1365 & 0.32\,ms & 3738 \\
resnet50             & 3.30\,ms & 238 & 2.34\,ms & 690 & 0.91\,ms & 1148 & 0.40\,ms & 2844 \\
resnet101            & 5.00\,ms & 174 & 3.63\,ms & 557 & 1.42\,ms &  714 & 0.69\,ms & 1554 \\
resnext101\_32x8d    & 10.50\,ms &  87 & 20.10\,ms & 270 & 3.39\,ms &  296 & 0.86\,ms & 1236 \\
tiny\_vit\_5m\_224   & 6.10\,ms & 133 & 16.70\,ms & 257 & 3.03\,ms &  334 & 0.71\,ms & 1515 \\
vit\_tiny\_p16\_224  & 3.90\,ms & 217 & 5.81\,ms & 420 & 1.52\,ms &  666 & 0.48\,ms & 2317 \\
\bottomrule
\end{tabularx}
\end{table*}

\subsection{Axelera Metis Multi-Chip Scaling}
\label{sec:metis_scaling}

The quad-chip Axelera Metis PCIe card (856\,TOPS peak) was tested on the subset of models that compiled successfully on the single-chip variant. As shown in Table~\ref{tab:metis_scaling}, throughput scales near-linearly across standard convolutional architectures. MNASNet-0.75 scales from 1,156\,FPS to 4,622\,FPS ($4.00\times$); MobileNetV4-Conv-S achieves the same $4.00\times$ factor. Heavier ResNet backbones range from $3.68\times$ (ResNet-101) to $4.60\times$ (ResNet-34), reflecting differences in workload-partitioning efficiency across model sizes. For workloads within Axelera's operator support envelope, the multi-core card represents a compelling path to very high throughput in a compact, low-power form factor.

\begin{table*}[t]
\centering
\caption{Axelera Metis 1$\times$ vs.\ 4$\times$ throughput scaling}
\label{tab:metis_scaling}
\small
\begin{tabularx}{\textwidth}{@{} X r r r r @{}}
\toprule
\textbf{Model} & \textbf{Lat.} & \textbf{FPS@1} & \textbf{FPS@4} & \textbf{Scale} \\
\midrule
mnasnet\_075         & 2.0\,ms &  1156 &  4622 & 4.00$\times$ \\
mobilenetv4\_conv\_s & 1.5\,ms &  1581 &  6331 & 4.00$\times$ \\
resnet18             & 1.8\,ms &  1131 &  4552 & 4.03$\times$ \\
resnet34             & 2.5\,ms &   580 &  2668 & 4.60$\times$ \\
resnet50             & 2.9\,ms &   516 &  2211 & 4.28$\times$ \\
resnet101            & 4.6\,ms &   330 &  1215 & 3.68$\times$ \\
\bottomrule
\end{tabularx}
\end{table*}

\subsection{Hailo-8 Multi-Stream Throughput}
\label{sec:hailo_streaming}

The Hailo-8 implements a pipelined dataflow architecture that partitions a neural network across fixed hardware stages. When six inference jobs are submitted concurrently, they overlap in pipeline fashion, yielding a near-$6\times$ increase in throughput at unchanged per-request latency. Table~\ref{tab:hailo_streaming} reports single-stream (FPS@1) and six-stream (FPS@6) figures for all models that compiled successfully. Throughput scales almost exactly $6\times$ for lightweight convolutional models; the reduced multipliers observed for ResNet-101 ($5.31\times$) and ResNeXt-101 ($5.48\times$) are a direct consequence of on-chip memory constraints discussed in Section~\ref{sec:discussion}.

\begin{table*}[t]
\centering
\caption{Hailo-8 single-stream vs.\ six-stream pipelined throughput}
\label{tab:hailo_streaming}
\small
\begin{tabularx}{\textwidth}{@{} X r r r r @{}}
\toprule
\textbf{Model} & \textbf{Lat.} & \textbf{FPS@1} & \textbf{FPS@6} & \textbf{Scale} \\
\midrule
mnasnet\_075          & 1.87\,ms &  3081 & 18485 & 6.00$\times$ \\
mobilenetv4\_conv\_s  & 1.28\,ms &  5948 & 35684 & 6.00$\times$ \\
resnet18              & 1.20\,ms &  2509 & 15053 & 6.00$\times$ \\
resnet34              & 4.02\,ms &  1139 &  6838 & 6.00$\times$ \\
resnet50              & 5.02\,ms &   771 &  4625 & 6.00$\times$ \\
resnet101             & 7.17\,ms &   112 &   595 & 5.31$\times$ \\
resnext101\_32x8d     & 15.17\,ms &   61 &   334 & 5.48$\times$ \\
vit\_tiny\_p16\_224   & 9.66\,ms &    94 &   567 & 6.03$\times$ \\
\bottomrule
\end{tabularx}
\end{table*}

\subsection{EdgeCortix Sakura~II Dual-Chip Throughput}
\label{sec:sakura_dualchip}

EdgeCortix Sakura~II was also evaluated in a dual-chip configuration (two chips partitioned together). Table~\ref{tab:sakura_dualchip} compares single-chip and dual-chip latency and throughput for all models that completed deployment in both configurations. Unlike the Axelera Metis, which scales near-linearly up to $4.60\times$ with four cores, the Sakura~II dual-chip configuration delivers a more modest throughput gain of $1.13$--$1.33\times$, suggesting that inter-chip communication or workload-partitioning overhead partially offsets the additional compute capacity.

It should be noted that the dual-chip partition mode evaluated here is not the only deployment option when two Sakura~II chips are available. Rather than joining both chips into a single larger partition for one network, each chip can instead be programmed independently to run a \emph{different} network. This enables two fully independent inference pipelines to run in parallel, each with single-chip throughput, effectively doubling the number of concurrent models that can be served. For applications such as multi-task vision systems or sensor-fusion pipelines that require separate networks running simultaneously, this independent partition mode may be more practical than the throughput scaling mode benchmarked above.

\begin{table*}[t]
\centering
\caption{EdgeCortix Sakura~II 1$\times$ vs.\ 2$\times$ chip throughput}
\label{tab:sakura_dualchip}
\small
\begin{tabularx}{\textwidth}{@{} X r r r r r @{}}
\toprule
\textbf{Model} & \multicolumn{2}{c}{\textbf{1$\times$}} & \multicolumn{2}{c}{\textbf{2$\times$}} & \textbf{Scale} \\
\cmidrule(lr){2-3}\cmidrule(lr){4-5}
 & \textbf{Lat.} & \textbf{FPS} & \textbf{Lat.} & \textbf{FPS} & \\
\midrule
mnasnet\_075          & 1.02\,ms &   956 & 0.99\,ms &  1270 & 1.33$\times$ \\
mobilenetv3\_small    & 0.83\,ms &  1162 & 0.70\,ms &  1459 & 1.26$\times$ \\
mobilenetv4\_conv\_s  & 0.76\,ms &  1262 & 0.65\,ms &  1461 & 1.16$\times$ \\
resnet18              & 1.09\,ms &   894 & 0.83\,ms &  1016 & 1.14$\times$ \\
resnet34              & 1.56\,ms &   630 & 1.08\,ms &   787 & 1.25$\times$ \\
resnet50              & 1.94\,ms &   507 & 1.49\,ms &   653 & 1.29$\times$ \\
resnet101             & 2.79\,ms &   355 & 2.12\,ms &   464 & 1.31$\times$ \\
resnext101\_32x8d     & 14.32\,ms &   71 & 12.67\,ms &    80 & 1.13$\times$ \\
\bottomrule
\end{tabularx}
\end{table*}

\subsection{NVIDIA A5000 Quantization Strategy}
\label{sec:nv_quant}

Tables~\ref{tab:nv_latency}--\ref{tab:nv_accuracy} break down the effect of quantization format on latency, throughput, and top-1 accuracy for the A5000 with TensorRT. Table~\ref{tab:op_coverage} reports the fraction of operations executed per precision for selected models, illustrating where hard versus soft calibration diverges.

The transition from FP32 to FP16 already provides substantial gains for convolutional architectures: ResNet-18 improves from 1,508 to 4,335\,FPS ($2.9\times$). Moving to INT8 provides a further step: ResNet-18 reaches 6,800\,FPS under INT8 Weak ($4.5\times$ over FP32). ModelOpt-assisted INT8 consistently outperforms both Weak and Hard INT8, delivering 20--60\% gains over standard INT8 for convolutional models.

Transformer architectures exhibit markedly different behaviour. For Tiny-ViT-5M and ViT-Tiny, INT8 Hard calibration \emph{increases} latency relative to INT8 Weak, because it forces 94.5\% and 96.5\% of operations to FP32 respectively (Table~\ref{tab:op_coverage}). Soft calibration retains attention operations in FP16 while achieving INT8 coverage for the remaining layers. ModelOpt recovers throughput and achieves the best result for transformers (2,615\,FPS for ViT-Tiny).

Accuracy losses are modest for convolutional architectures ($<$0.5\% top-1 degradation for ResNet/ResNeXt). Notable outliers are EfficientNet-B0 ($\approx$3.9 percentage-point loss under INT8 Weak) and MobileNetV3-Small, which loses 13.8 points under INT8 Hard and a severe 53 points under ModelOpt INT8—a warning that depthwise-heavy architectures with squeeze-and-excitation blocks may require mixed-precision strategies.

\begin{table*}[t]
\centering
\caption{Latency (ms) by quantization format on NVIDIA RTX A5000 with TensorRT}
\label{tab:nv_latency}
\small
\begin{tabularx}{\textwidth}{@{} X r r r r r @{}}
\toprule
\textbf{Model} & \textbf{FP32} & \textbf{FP16} & \textbf{INT8 Weak} & \textbf{INT8 Hard} & \textbf{INT8 ModelOpt} \\
\midrule
convnext\_tiny       & 1.23 & 0.87 & 0.87 & 1.09 & 0.76 \\
efficientnet\_b0     & 0.78 & 0.64 & 0.57 & 0.57 & 0.40 \\
mnasnet\_075         & 0.36 & 0.29 & 0.27 & 0.27 & 0.16 \\
mobilenetv3\_small   & 0.41 & 0.34 & 0.34 & 0.35 & 0.28 \\
mobilenetv4\_conv\_s & 0.32 & 0.24 & 0.24 & 0.24 & 0.13 \\
resnet18             & 0.71 & 0.28 & 0.20 & 0.20 & 0.12 \\
resnet34             & 1.32 & 0.47 & 0.32 & 0.31 & 0.22 \\
resnet50             & 1.22 & 0.53 & 0.40 & 0.40 & 0.27 \\
resnet101            & 2.27 & 0.95 & 0.69 & 0.69 & 0.51 \\
resnext101\_32x8d    & 2.52 & 1.20 & 0.86 & 0.86 & 0.67 \\
tiny\_vit\_5m\_224   & 1.00 & 0.70 & 0.71 & 1.21 & 0.75 \\
vit\_tiny\_p16\_224  & 0.76 & 0.48 & 0.48 & 1.09 & 0.38 \\
\bottomrule
\end{tabularx}
\end{table*}

\begin{table*}[t]
\centering
\caption{Throughput (FPS) by quantization format on NVIDIA RTX A5000 with TensorRT}
\label{tab:nv_fps}
\small
\begin{tabularx}{\textwidth}{@{} X r r r r r @{}}
\toprule
\textbf{Model} & \textbf{FP32} & \textbf{FP16} & \textbf{INT8 Weak} & \textbf{INT8 Hard} & \textbf{INT8 ModelOpt} \\
\midrule
convnext\_tiny       & 846  & 1219 & 1218 &  959 & 1298 \\
efficientnet\_b0     & 1356 & 1692 & 1898 & 1918 & 2450 \\
mnasnet\_075         & 3229 & 4138 & 4481 & 4466 & 5895 \\
mobilenetv3\_small   & 2734 & 3450 & 3393 & 3318 & 3503 \\
mobilenetv4\_conv\_s & 3775 & 5170 & 5299 & 5227 & 7221 \\
resnet18             & 1508 & 4335 & 6800 & 6739 & 8236 \\
resnet34             &  790 & 2354 & 3738 & 3763 & 4519 \\
resnet50             &  854 & 2079 & 2844 & 2855 & 3631 \\
resnet101            &  450 & 1108 & 1554 & 1550 & 1929 \\
resnext101\_32x8d    &  405 &  867 & 1236 & 1238 & 1468 \\
tiny\_vit\_5m\_224   & 1045 & 1529 & 1515 &  865 & 1314 \\
vit\_tiny\_p16\_224  & 1409 & 2301 & 2317 &  964 & 2615 \\
\bottomrule
\end{tabularx}
\end{table*}

\begin{table*}[t]
\centering
\caption{Top-1 accuracy (\%) by quantization format. ONNX column shows the floating-point reference.}
\label{tab:nv_accuracy}
\small
\begin{tabularx}{\textwidth}{@{} X r r r r r r @{}}
\toprule
\textbf{Model} & \textbf{ONNX ref.} & \textbf{FP32} & \textbf{FP16} & \textbf{INT8 Weak} & \textbf{INT8 Hard} & \textbf{INT8 ModelOpt} \\
\midrule
convnext\_tiny       & 55.28 & 55.28 & 55.31 & 55.18 & 55.59 & 55.27 \\
efficientnet\_b0     & 95.23 & 95.23 & 95.28 & 91.36 & 91.36 & 83.78 \\
mnasnet\_075         & 66.21 & 66.26 & 66.19 & 65.96 & 65.96 & 67.18 \\
mobilenetv3\_small   & 86.06 & 86.06 & 86.03 & 85.29 & 72.22 & 32.81 \\
mobilenetv4\_conv\_s & 86.19 & 86.24 & 86.16 & 86.98 & 86.98 & 87.59 \\
resnet18             & 94.87 & 94.87 & 94.87 & 94.98 & 94.98 & 94.72 \\
resnet34             & 95.94 & 95.94 & 96.00 & 95.87 & 95.87 & 95.99 \\
resnet50             & 97.37 & 97.37 & 97.35 & 97.24 & 97.24 & 97.07 \\
resnet101            & 96.68 & 96.68 & 96.71 & 96.73 & 96.73 & 96.48 \\
resnext101\_32x8d    & 88.73 & 88.73 & 88.76 & 88.38 & 88.35 & 89.45 \\
tiny\_vit\_5m\_224   & 94.16 & 94.14 & 94.16 & 90.98 & 90.85 & 90.72 \\
vit\_tiny\_p16\_224  & 80.02 & 80.02 & 80.02 & 80.02 & 79.97 & 79.10 \\
\bottomrule
\end{tabularx}
\end{table*}

\begin{table*}[t]
\centering
\caption{Fraction of operations executed per precision format (selected models)}
\label{tab:op_coverage}
\small
\begin{tabularx}{\textwidth}{@{} X r r r @{}}
\toprule
\textbf{Model + Format} & \textbf{FP32} & \textbf{FP16} & \textbf{INT8} \\
\midrule
convnext\_tiny – FP32    & 100.0\% & 0.0\%  & 0.0\%  \\
convnext\_tiny – FP16    &   4.7\% & 95.3\% & 0.0\%  \\
convnext\_tiny – INT8 Weak &  4.6\% & 94.6\% & 0.8\%  \\
convnext\_tiny – INT8 Hard & 72.5\% &  0.0\% & 27.5\% \\
\midrule
efficientnet\_b0 – FP32  & 100.0\% & 0.0\%  & 0.0\%  \\
efficientnet\_b0 – FP16  &   0.5\% & 99.5\% & 0.0\%  \\
efficientnet\_b0 – INT8 Weak & 1.7\% & 0.0\% & 98.3\% \\
efficientnet\_b0 – INT8 Hard & 1.7\% & 0.0\% & 98.3\% \\
\midrule
resnet18 – FP32          & 100.0\% & 0.0\%  & 0.0\%  \\
resnet18 – FP16          &   4.2\% & 95.8\% & 0.0\%  \\
resnet18 – INT8 Weak     &   8.0\% &  0.0\% & 92.0\% \\
resnet18 – INT8 Hard     &   8.0\% &  0.0\% & 92.0\% \\
\midrule
tiny\_vit – FP32         & 100.0\% & 0.0\%  & 0.0\%  \\
tiny\_vit – FP16         &   1.2\% & 98.8\% & 0.0\%  \\
tiny\_vit – INT8 Weak    &   1.2\% & 89.0\% & 9.8\%  \\
tiny\_vit – INT8 Hard    &  94.5\% &  0.0\% & 5.5\%  \\
\midrule
vit\_tiny – FP32         & 100.0\% & 0.0\%  & 0.0\%  \\
vit\_tiny – FP16         &   2.2\% & 97.8\% & 0.0\%  \\
vit\_tiny – INT8 Weak    &   2.2\% & 97.8\% & 0.0\%  \\
vit\_tiny – INT8 Hard    &  96.5\% &  0.0\% & 3.5\%  \\
\bottomrule
\end{tabularx}
\end{table*}

\subsection{Discussion}
\label{sec:discussion}

\subsubsection{Hailo Pipelining vs.\ DSP Sequential Execution}

The Hailo-8 implements a pipelined dataflow architecture in which a neural network is partitioned into stages assigned to dedicated hardware blocks. Multiple inference jobs overlap in pipeline fashion, analogous to CPU instruction pipelining at the inference level. The practical consequence is illustrated directly in Table~\ref{tab:hailo_streaming}: for MNASNet-0.75, single-stream throughput is 3,081\,FPS, while six-stream pipelined throughput reaches 18,485\,FPS—a $6\times$ multiplier that closely tracks the number of concurrent pipeline stages.

The Qualcomm Hexagon DSP operates sequentially: the entire DSP fabric is occupied by a single inference job from input to output, and a second job cannot begin until the first has completed. No comparable scaling is achievable regardless of concurrent requests. This distinction has major implications for video analytics pipelines or batch inference services prioritizing aggregate throughput.

\subsubsection{Single-Stream Latency}

When only single-stream latency is considered, the picture reverses. The A5000 consistently achieves the lowest latency, often 3--10$\times$ lower than the ASIC NPUs. For ResNet-18, the A5000 achieves 0.20\,ms versus 1.20\,ms (Hailo-8) and 1.8\,ms (Axelera Metis). The Qualcomm IQ-9075 occupies a useful middle ground, reaching 0.50\,ms for ResNet-18 and approaching GPU-class latency for lightweight models. The Hailo-8's relatively high single-stream latency is a direct consequence of its pipelined architecture: a single request must traverse the full pipeline depth before producing output.

\subsubsection{Platform-Level Takeaways}

\textbf{NVIDIA (A5000 + TensorRT).} Full model coverage across all quantization configurations, highest single-stream throughput, and a mature toolchain. Real-world multi-stream throughput benefits are more limited than the raw TOPS figure suggests, as batch scheduling overhead and memory bandwidth constraints become bottlenecks at high concurrency.

\textbf{Qualcomm (QCS6490, QCS8550, IQ-9075).} Near-complete model coverage and competitive single-stream latency; the IQ-9075 approaches GPU-class latency for lightweight models. Sequential execution prevents pipelined throughput scaling. The SDK ecosystem is functional but fragmented, with multiple SDK generations (SNPE, QNN, AI Engine Direct) adding integration overhead.

\textbf{Hailo-8.} Outstanding multi-stream throughput within its operator support envelope, with FPS@6 figures competitive with or exceeding the A5000 for several lightweight convolutional models. The compiler and runtime APIs are notably well-designed. The primary limitation is operator coverage: four of twelve benchmark models could not be compiled. A discrepancy was observed between our measured throughput for ResNet-34 and ResNet-50 and the figures reported in official Hailo benchmarks~\cite{ref26}. We attribute this to an architectural difference between the PyTorch model weights used in this evaluation (sourced from \texttt{torchvision}) and the TensorFlow-derived variants used in Hailo's published results: the TensorFlow variants appear to reduce channel width more aggressively in early stages, resulting in a lighter compute graph. The exact reason of the difference is not fully confirmed, but readers should keep this in mind when comparing these specific figures against vendor-published numbers.

A further constraint is the Hailo-8's fixed on-chip SRAM budget. The device carries no external DRAM; all activations and weights must reside entirely on-chip. Larger models such as ResNet-101 and ResNeXt-101 exceed this budget and must be partitioned by the compiler into multiple execution contexts, each of which is loaded and dispatched sequentially. This sequential multi-context execution breaks the assumptions of the pipelined dataflow architecture: rather than a single continuous pipeline stage, the hardware processes several independent sub-graphs in sequence, stalling between them. The performance penalty is clearly visible in Tables~\ref{tab:asic_results} and~\ref{tab:hailo_streaming}: while lightweight models achieve near-$6\times$ six-stream scaling, ResNet-101 and ResNeXt-101 reach only $5.31\times$ and $5.48\times$ respectively. The same constraint also depresses single-stream throughput for these models—ResNet-101 reports 112\,FPS compared to 771\,FPS for ResNet-50, a drop disproportionate to the $1.7\times$ difference in parameter count. This on-chip memory ceiling is therefore an important practical consideration when targeting the Hailo-8 with deeper or wider architectures.

\textbf{Axelera Metis.} Highest peak TOPS among single-chip ASIC NPUs evaluated, with near-linear multi-chip throughput scaling (Section~\ref{sec:metis_scaling}). Operator coverage mirrors Hailo-8 and represents the same fundamental constraint.

\textbf{EdgeCortix Sakura~II.} Partially different operator coverage from Hailo and Axelera, succeeding on MobileNetV3 where both ASIC competitors failed. Dual-chip throughput scaling is moderate ($1.13$--$1.33\times$, Table~\ref{tab:sakura_dualchip}), lower than the near-linear scaling of Axelera Metis, but BF16 support and larger on-board DRAM make it more flexible for mixed-precision and generative AI workloads.

\subsubsection{Optimisation Takeaways}

\textbf{INT8 quantization} is effectively a solved problem for standard convolutional architectures: PTQ with appropriate calibration recovers near-lossless accuracy while delivering 2--5$\times$ throughput improvements over FP32. For transformer-based architectures, soft calibration (INT8 Weak) is preferable over aggressive hard calibration, as shown in Table~\ref{tab:op_coverage}.

\textbf{Sparsification} without dedicated hardware support yields negligible computational savings. On the NVIDIA platform, where 2:4 structured sparsity is hardware-accelerated, meaningful throughput improvements are achievable for convolutional models.

\textbf{Pruning} delivered modest improvements unless the model was substantially oversized relative to the task. Knowledge distillation consistently outperforms magnitude-based pruning as a model compression strategy when a smaller target architecture is acceptable.

\textbf{Network customisation}: adapting the operator mix to the specific constraints and strengths of the target hardware remains the highest-leverage optimization strategy for dedicated NPUs. Replacing operator types outside the ASIC's supported set, or restructuring layers to maximize pipeline utilization on the Hailo-8, can mean the difference between a \FAIL{} and a fully functional high-throughput deployment.

% ============================================================
\clearpage
\balance
\section*{References}

\begingroup
\small
\setlength{\parskip}{2pt}
\setlength{\itemsep}{0pt}

\endgroup


\begin{thebibliography}{99}

\bibitem{ref1} SNS Insider. ``AI Inference Market to Reach USD 349.49 Billion by 2032.'' GlobeNewswire, Sept.\ 24, 2025. \url{https://www.globenewswire.com/news-release/2025/09/24/3155215/0/en/AI-Inference-Market-to-Reach-USD-349-49-Billion-by-2032}

\bibitem{ref2} VentureBeat. ``5\% GPU utilization: The \$401 billion AI infrastructure problem enterprises can't keep ignoring.'' May 2026. \url{https://venturebeat.com/infrastructure/5-gpu-utilization-the-401-billion-ai-infrastructure-problem-enterprises-cant-keep-ignoring}

\bibitem{ref3} Kim et al.\ ``Performance and Efficiency Gains of NPU-Based Servers over GPUs for AI Model Inference.'' \textit{MDPI Systems} 13(9):797, Sept.\ 11, 2025. \url{https://www.mdpi.com/2079-8954/13/9/797}

\bibitem{ref4} IntuitionLabs. ``LLM Inference Hardware: An Enterprise Guide to Key Players.'' Mar.\ 1, 2026. \url{https://intuitionlabs.ai/articles/llm-inference-hardware-enterprise-guide}

\bibitem{ref5} Kneron / Autonomy Global. ``Kneron Unveils Enterprise-Grade On-Premises AI Infrastructure Platform.'' Jan.\ 8, 2026. \url{https://www.autonomyglobal.co/kneron-unveils-enterprise-grade-on-premises-ai-infrastructure-platform/}

\bibitem{ref6} Persistence Market Research. ``AI Laptop Market Share.'' 2025. \url{https://www.persistencemarketresearch.com/market-research/ai-laptop-market.asp}

\bibitem{ref7} Equinix Blog. ``Data Sovereignty and AI: Why You Need Distributed Infrastructure.'' May 2025. \url{https://blog.equinix.com/blog/2025/05/14/data-sovereignty-and-ai-why-you-need-distributed-infrastructure/}

\bibitem{ref8} MarketsandMarkets. ``AI Inference Market Size, Share \& Growth, 2025 to 2030.'' \url{https://www.marketsandmarkets.com/Market-Reports/ai-inference-market-189921964.html}

\bibitem{ref9} Contabo. ``NPU vs GPU: Differences in AI Processing.'' Feb.\ 4, 2026. \url{https://contabo.com/blog/npu-vs-gpu/}

\bibitem{ref10} OrdinaryTech. ``On-Device AI in 2026: How NPUs Are Transforming AI PCs.'' Mar.\ 5, 2026. \url{https://ordinarytech.ca/blogs/news/on-device-ai-in-2026-how-npus-are-transforming-ai-pcs-for-creators-and-power-users-1}

\bibitem{ref11} ArticSledge / WiseGuy Reports. ``What is a Neural Processing Unit (NPU)? Complete 2026 Guide.'' Apr.\ 19, 2026. \url{https://www.articsledge.com/post/neural-processing-unit-npu}

\bibitem{ref12} MarketReportsWorld. ``Network Processing Unit (NPU) Market Share \& Trends [2033].'' Mar.\ 2026. \url{https://www.marketreportsworld.com/market-reports/network-processing-unit-npu-market-14720926}

\bibitem{ref13} OxMaint. ``AI Data Sovereignty Compliance Guide for Regulated Industries.'' May 2, 2026. \url{https://oxmaint.com/sap-integration/on-prem-ai/ai-data-sovereignty-compliance}

\bibitem{ref14} AMD. ``AMD Ryzen AI PRO Processor Leadership and TCO Benefits.'' 2025. \url{https://www.amd.com/en/blogs/2025/amd-ryzen-ai-pro-processor-leadership-and-tco-benefits}

\bibitem{ref15} R.\ Wightman. ``PyTorch Image Models (timm).'' HuggingFace, 2019. \url{https://github.com/huggingface/pytorch-image-models}

\bibitem{ref16} Axelera AI. ``Europa AIPU.'' Axelera AI, 2026. \url{https://axelera.ai/ai-accelerators/aipu/europa}

\bibitem{ref17} Hailo. ``Hailo-8 / Hailo-10 AI Accelerator Product Brief.'' Hailo Technologies, 2024. \url{https://hailo.ai/products/ai-accelerators/}

\bibitem{ref26} Hailo. ``Hailo Model Zoo --- Hailo-8 Classification Models.'' GitHub, 2024. \url{https://github.com/hailo-ai/hailo_model_zoo/blob/master/docs/public_models/HAILO8/HAILO8_classification.rst}

\bibitem{ref18} Axelera AI. ``Metis PCIe AI Inference Acceleration Cards.'' Axelera AI, 2024. \url{https://axelera.ai/ai-accelerators/metis-pcie-ai-acceleration-card}

\bibitem{ref19} EdgeCortix. ``Sakura-II AI Accelerator Datasheet.'' EdgeCortix Inc., 2024. \url{https://www.edgecortix.com/en/products/sakura}

\bibitem{ref20} SiMa.ai. ``MLSoC Product Brief.'' SiMa.ai, 2024. \url{https://sima.ai/products/mlsoc/}

\bibitem{ref21} Qualcomm. ``QCS6490 / QCS8550 Product Pages.'' Qualcomm Technologies, 2024. \url{https://www.qualcomm.com/products/internet-of-things/industrial/building-enterprise/qcs6490}, \url{https://www.qualcomm.com/internet-of-things/products/q8-series/qcs8550}

\bibitem{ref25} Qualcomm. ``Dragonwing IQ9 Series Product Brief.'' Qualcomm Technologies, 2026. \url{https://docs.qualcomm.com/doc/87-83840-1/87-83840-1_REV_A_Qualcomm_IQ9_Series_Product_Brief.pdf}

\bibitem{ref22} Intel. ``Intel Core Ultra 200V Series Processors (Lunar Lake).'' Intel Corporation, 2024. \url{https://www.intel.com/content/www/us/en/ark/products/series/236800/intel-core-ultra-200v-series-processors.html}

\bibitem{ref23} NVIDIA. ``NVIDIA RTX A5000 Datasheet.'' NVIDIA Corporation, 2021. \url{https://www.nvidia.com/content/dam/en-zz/Solutions/products/workstations/nvidia-rtx-a5000-datasheet.pdf}

\bibitem{ref24} NVIDIA. ``NVIDIA Ampere Architecture In-Depth.'' NVIDIA Developer Blog, 2020. \url{https://developer.nvidia.com/blog/nvidia-ampere-architecture-in-depth/}

\end{thebibliography}
\end{document}